\documentclass[%
reprint,
superscriptaddress,
 amsmath,amssymb,
 aps,
showpacs,
]{revtex4-2}

\DeclareUnicodeCharacter{0301}{\'{e}}

\usepackage{sistyle}
\usepackage{graphicx}
\usepackage{dcolumn}
\usepackage{soul}
\usepackage[usenames,dvipsnames]{color}
\usepackage{bm}

\newcommand{\SB}[1]{\textit{\emph{\textcolor{black}{#1}}}}

\newcommand{\D}{{\mathrm d}}

\begin{document}

\preprint{APS/123-QED}


\title{Experimental mitigation of fast magnetic reconnection in multiple interacting laser-produced plasmas}

\author{S. Bola\~{n}os }
\email{simon.bolanos@protonmail.ch}
\affiliation{LULI, CNRS, CEA, Sorbonne Université, Ecole Polytechnique, Institut Polytechnique de Paris - F-91128 Palaiseau cedex, France}
\affiliation{LPP, Sorbonne Université, CNRS, Ecole Polytechnique, F-91128 Palaiseau, France}
\author{R. Smets}
\affiliation{LPP, Sorbonne Université, CNRS, Ecole Polytechnique, F-91128 Palaiseau, France}
\author{C. Courtois}
\affiliation{CEA, DAM, DIF, F-91297 Arpajon, France}
\author{N. Blanchot}
\affiliation{CEA, DAM, CESTA, F-33114 Le Barp, France}
\author{G. Boutoux}
\affiliation{CEA, DAM, CESTA, F-33114 Le Barp, France}
\author{W. Cayzac}
\affiliation{CEA, DAM, DIF, F-91297 Arpajon, France}
\author{S. N. Chen}
\affiliation{IFIN-HH, ”Horia Hulubei” National Institute of Physics and Nuclear Engineering, Bucharest - Magurele, Romania}
\author{V. Denis}
\affiliation{CEA, DAM, CESTA, F-33114 Le Barp, France}
\author{A. Grisollet}
\affiliation{CEA, DAM, DIF, F-91297 Arpajon, France}
\author{I. Lantuejoul}
\affiliation{CEA, DAM, DIF, F-91297 Arpajon, France}
\author{L. Le Deroff}
\affiliation{CEA, DAM, CESTA, F-33114 Le Barp, France}
\author{R. Riquier}
\affiliation{CEA, DAM, DIF, F-91297 Arpajon, France}
\author{B. Vauzour}
\affiliation{CEA, DAM, DIF, F-91297 Arpajon, France}
\author{J. Fuchs}
\email{julien.fuchs@polytechnique.edu}
\affiliation{LULI, CNRS, CEA, Sorbonne Université, Ecole Polytechnique, Institut Polytechnique de Paris - F-91128 Palaiseau cedex, France}

\date{\today}

\begin{abstract}

The meeting of astrophysical plasmas and their magnetic fields creates many reconnection sites. We experimentally compare the reconnection rate of laser-driven magnetic reconnection when it takes place at a single site and multiple sites. For a single site, where the ram pressure dominates the magnetic pressure, the measured reconnection rate exceeds the well-established rate of 0.1. However, in the case of multiple close-by sites, we observed a reduction of the reconnection rate. Hybrid-PIC simulations support this observation and suggest that the distortion of the Hall field as well as the concomitant obstruction of one of the outflows are detrimental to the magnetic reconnection rate.

\end{abstract}


\maketitle

Magnetic reconnection (MR) is the universal process by which, when two anti-parallel magnetic field-lines in a plasma meet, the magnetic energy stored in the associated current sheet is converted in a combination of kinetic and thermal plasma energy~\cite{parker1957,sweet1958,petschek1964}.
Understanding the dynamics of MR in detail is crucial, since it is suspected to be the source for turbulence~\cite{Loureiro2017} and particle acceleration~\cite{Adhikari2019} in a large variety of astrophysical phenomena~\cite{Drake2010,Guo2015,Werner2017}.

\begin{figure}[ht!]
\includegraphics[width=0.4\textwidth]{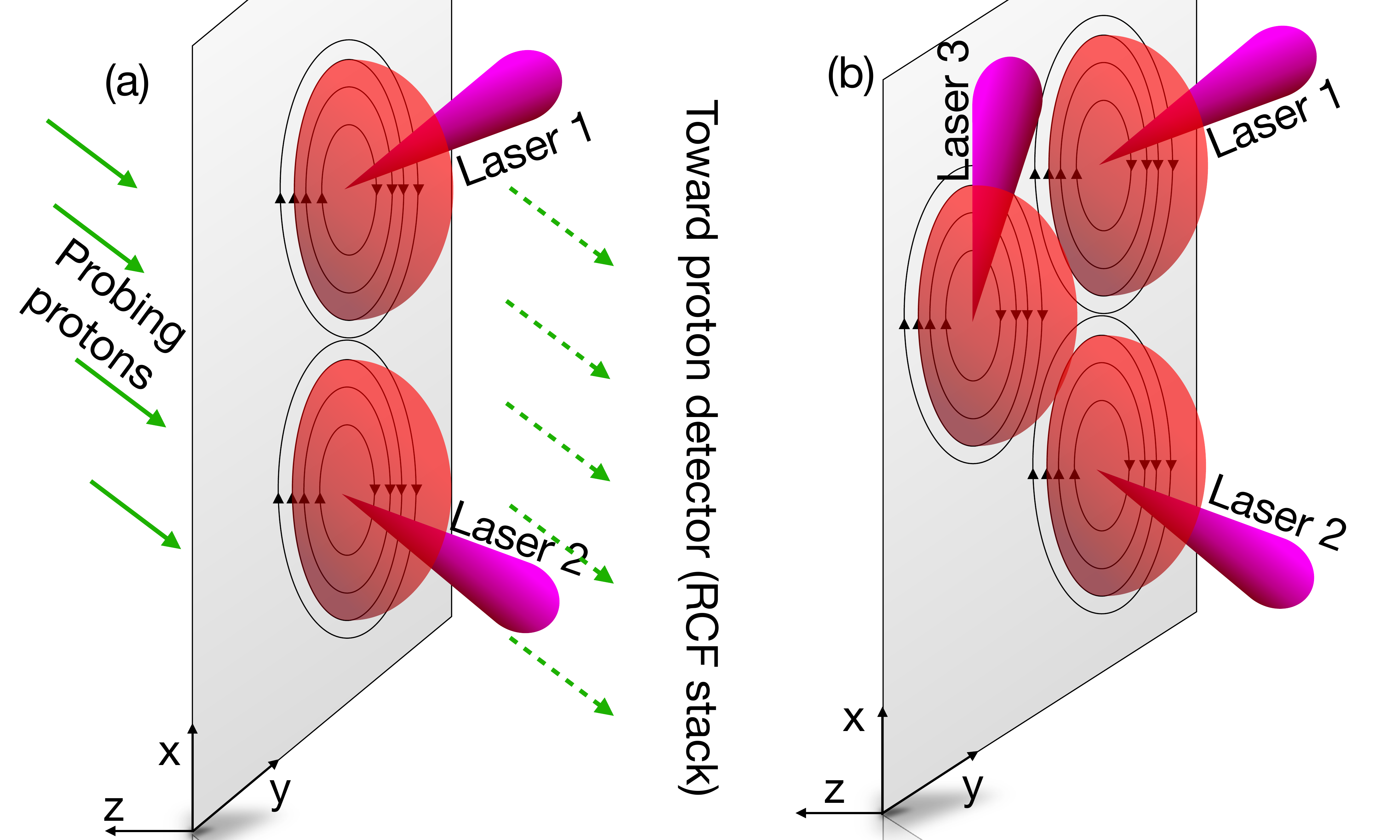}
\caption{\label{fig:setup}   3D cartoon 
of the experiment 
performed at the LMJ laser facility to investigate magnetic reconnection between two or three adjacent, laser-driven magnetic field structures (illustrated by the dark arrows). Panel (a-b) illustrate, respectively, the two- and three-plume cases investigated here.
The green arrows illustrate the 
 proton beam utilized for 
probing the magnetic fields in the plasmas and their dynamics following reconnection.}
\end{figure}

In a simplified two-dimensional MR topology (see Fig.\ref{fig:setup}(a)), the plasma inflows move along the $x$-axis while the unperturbed magnetic fields are oriented along the $y$-axis, and the efficiency at which the magnetic flux is converted into plasma energy is controlled by the $z$-component of the electric field.
This out-of-plane component of the electric field,
also called the reconnection rate~\cite{sweet1958,vasyliunas1975}, as computed by numerical simulations (Hall-MHD, hybrid-PIC and full-PIC codes) and inferred from observations, has been estimated to be close to 0.1~\cite{cassak2017} for fast reconnection~\cite{birn2001+}, although no clear explanations as to why this rate is $\sim$0.1 so has yet emerged.

Aside from direct in-situ measurements of MR by spacecrafts in the magnetosphere~\cite{ergun2016,phan2018},  laboratory experiments have been designed to investigate the microphysics associated with MR~\cite{yamada1997,egedal2007,gekelman2012,fox2018}.
Among these, high-power lasers, complemented by numerical simulations~\cite{fox2011,smets2014}, are an appropriate tool, since they allow creating two magnetized expanding plasmas (see Fig.\ref{fig:setup}(a)), as well as access to the mapping of the magnetic field, and measure the plasma density and temperature ~\cite{nilson2006,Li2007}.

In this Letter, we explore a deviation from the idealized picture of MR\cite{nilson2006,Li2007} where we concomitantly trigger MR at several sites, using high-power laser generated plasmas (see Fig.\ref{fig:setup}(b)).
Note that one previous work investigated magnetic fields generated  simultaneous at several sites~\cite{Li2007}, but these were too far from each other to magnetically interact.
Here, we purposely position the sites close-by, so that MR can take place simultaneously at multiple locations.
Note also that MR in our conditions is not in the plasmoid regime~\cite{loureiro2012}, but rather in the fast X-line regime. This is justified since, using hydro-radiative simulations performed with the code FCI2~~\cite{Dattolo2001}, we estimate the Lundqvist number~\cite{loureiro2012} to be $S \sim 10^3$ and that we have an aspect ratio of the current sheet, i.e. its length divided by its thickness, smaller than 50.

What we show here is that, depending on the number of reconnection sites, the balance, between the Biermann-battery effect~\cite{Biermann1950,Matteucci2018} that keeps creating magnetic flux and the MR that drains this magnetic flux out of the current sheet, is modified.
We quantitatively estimate the MR rate in the case of two reconnection sites, which proves to exceed the well-established rate of 0.1. In the case of three reconnection sites, MR is observed to be significantly slowed down. Following Ref.~\cite{smets2014}, our hybrid-PIC simulations suggest that this reduction is linked to the destabilization of the out-of-plane magnetic field that is associated with the current sheet compression.
These results do not only demonstrate that the MR rate can be significantly impacted by the topology of the involved magnetic fields, but they also shed light on the complexity of MR when many sites are involved, as is the case in many events occurring in the solar system ~\cite{Burlaga2002} or in black hole environment ~\cite{Bransgrove2021}.


The experiment was carried out at the Laser Megajoule (LMJ) facility (France)~\cite{Casner2015,Denis2021}.
We use two (Fig.\ref{fig:setup}(a)) or three (Fig.\ref{fig:setup}(b)) lasers to drive plasma plumes, i.e. spots where the laser energy is deposited on a solid target (5 $\mu$m-thick gold foils) and drives the generation of magnetic field through the Biermann-battery effect.
The two-plume case corresponds to the geometry of the canonical MR picture while in the three-plume case, MR occurs simultaneously at three sites arranged along a triangle.
The lasers were operating at $\lambda = \SI{351}{nm} $ wavelength, with a 5-ns square pulse duration and focal spots comprised between 240 and $\SI{ 350}{\micro m}$  FWHM diameter. Each laser beam had an energy of 12 kJ, yielding an on-target intensity of $\SI{4.8e14}{W/cm^2}$.

\begin{figure}[ht!]
\includegraphics[width=0.47\textwidth]{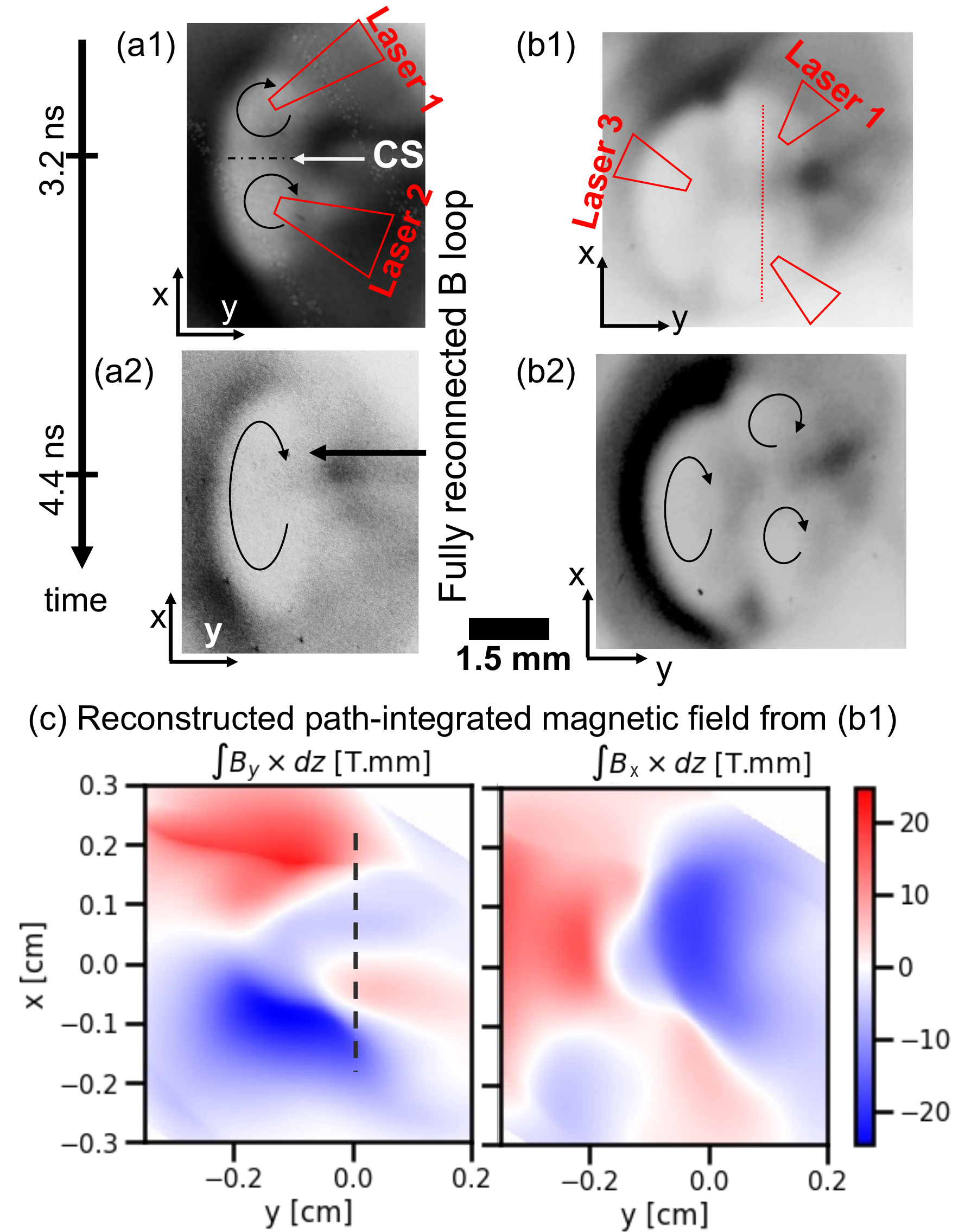}
\caption{\label{fig:rcf} (a1)-(a2) Timeline of  proton radiography images (measured using 23-MeV protons) where only two laser beams (depicted by the red cones in panel (a1)) were incident on target, separated by 1.15 mm. t=0 refers to the beginning of the lasers  irradiation on target. 
 Panels (b1)-(b2) are akin to  panels  (a1)-(a2), but for the three-plume case. The spatial scale  applies to all panels (a1-b2). Panel (c) displays the reconstructed path-integrated magnetic field of the proton image (b1), see text for details.  %
 }
\end{figure}

The primary diagnostic was proton deflectometry~\cite{reviewSchaeffer}. It employed as backlighter a high-energy (42-MeV maximum energy) proton beam produced
~\cite{Wilks2001} from a 25 µm thick gold foil by the high-intensity short pulse PETAL laser~\cite{Raffestin2021} (see the green arrows in Figure \ref{fig:setup}). As the protons travel across the plasmas, they are deflected by the magnetic fields; it is these deflections that induce proton beam dose modulations on the films onto which they are collected after crossing the plasmas. Several shots were performed by changing the delay at which the protons probed the plasma with respect to the start of the irradiation of the target by the main laser beams (see the times indicated in Fig.\ref{fig:rcf}(a1-a2)).  It is from the observed modulations on the films, and the changes over time of these modulations, that the magnetic field topology on the target and its evolution, are inferred~\cite{reviewSchaeffer}.

The magnetic field produced by the Biermann-battery mechanism is clockwise-oriented around the main plasma expansion axis (see Fig.\ref{fig:setup}) and therefore, the probing protons are radially deflected  outward from the plume~\cite{Cecchetti2009,Petrasso2009}. For most of the surface of the plasma plume, the magnetic field is compressed toward the target by the Nernst effect~\cite{Lancia2014}. However, at the edge of the plasma plumes, the magnetic field is radially advected with the plasma flow  and also expands away from the target (along the $z$-axis)~\cite{Campbell2020,Bolanos2022}. Due to geometrical constraints in the experiment, and as shown schematically in Fig.\ref{fig:setup}, the protons probe the plasma with an 34° angle with respect to the normal of the target surface. Therefore, the magnetic field in each plume appears elliptical in the proton images.

Figure \ref{fig:rcf} shows the proton images measured for the two cases investigated here. The whiter region on the proton images corresponds to a depletion of protons, after they have been deflected in the plasma by the Biermann-battery generated magnetic field. Conversely, the darker regions correspond to accumulations of protons deflected from surrounding regions. We can observe, around each laser spot in Fig.\ref{fig:rcf}, a depletion area surrounded by an accumulation of protons (darker rim). This corresponds to what is expected for such Biermann-battery magnetic structure~\cite{Petrasso2009,Cecchetti2009,Lancia2014}, that is, the strongest magnetic field is observed in the core of each plume, while it decrease at their outermost border.

In the case where we have only two plasma plumes interacting, we can observe at the earlier time (3.2 ns), around each laser spot, a dark proton ring (see Fig.\ref{fig:rcf}(a1)), i.e. corresponding to the juxtaposition of two Biermann-battery magnetic structures separated by a current sheet (CS). In the center of the current sheet,
the accumulation of proton reduces significantly compared to that observed in the outer part of the plasma plumes, consistently with the expected local annihilation of magnetic flux induced by MR~\cite{nilson2006,Li2007}.
At the latest stage (see Fig.\ref{fig:rcf}(a2)), the contrast of the proton flux at the current sheet drops substantially until being barely distinguishable, again consistently with the fact that MR has led to the full annihilation of the magnetic flux, and thus to a global topological rearrangement of the magnetic structure into one large magnetic structure that encompasses the outer edge of the two laser irradiation spots, as schematically delineated in Fig.\ref{fig:rcf}(a2).

In contrast, in the case where  three plasma plumes interact, we observe in Fig.\ref{fig:rcf}(b1-b2)) that  the accumulation of the proton at the current sheet is and remains stronger over time. In particular there is no global rearrangement in the magnetic structure as the one that takes place for the two sites interaction; here, even at late time, we can still observe the presence of three magnetic structures.

It is possible to reconstruct maps of 
the path-integrated magnetic field from the recorded proton images such as those shown in Fig.\ref{fig:rcf} 
~\cite{Bott2017,Graziani2017}. Fig.~\ref{fig:rcf}(c) display the reconstructed magnetic field for 30-MeV protons using the  {\sc problem} solver ~\cite{Bott2017,Graziani2017} for the three-plume case probed at 3.2 ns.

\begin{figure*}
\includegraphics[width=0.98\textwidth]{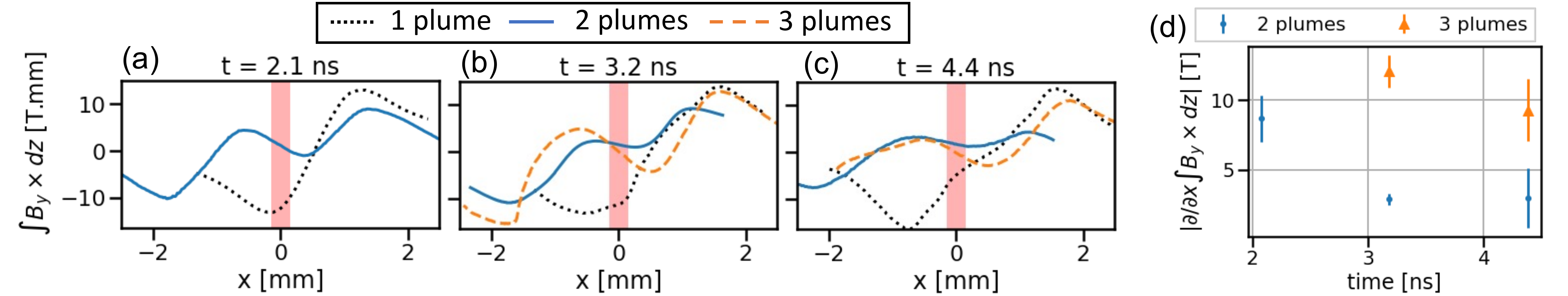}
\caption{\label{fig:B-inv} (a-c) Lineouts of the reconstructed path-integrated magnetic field at the reconnection site  for the three cases investigated here (employing respectively, a single laser beam, two laser beams, and three laser beams),  at three different times. For the three-plume case, the lineout is measured along the dashed line shown in  Fig~\ref{fig:rcf}(b1). Panel (d) summarizes the temporal evolution of the gradient of the path-integrated magnetic field at the current sheet location ($x=0$) for the two and three-plume  cases. Note that for the three-plume case, we  the same measurement performed at the two other reconnection sites are very consistent and lie in the depicted uncertainties.  }
\end{figure*}

The  expected magnetic field inversion at the current sheet can be observed in the path-integrated magnetic field maps. One of such inversion for the $B_y$ component can be seen in Fig.\ref{fig:rcf}(c1) at the location ($y=0$, $x=0$); it is quantitatively plotted in Fig.~\ref{fig:B-inv} where panels (a-c) display the spatial profile of the path-integrated $B_y$ magnetic field, for three probing times and for the two cases investigated here.
Fig.~\ref{fig:B-inv}(d) summarizes the temporal evolution of the $x$-gradient of the path integrated $B_y$ magnetic field, for the two (solid blue) and three (dashed orange) plumes cases.
For the sake of comparison, the magnetic field associated with a single plume (recorded in other shots during the same experiment), i.e. which is magnetically isolated and freely expanding, is also plotted (black dashed line). The strongest inversion is observed early on (at $t = 2.1$ ns, see Fig.~\ref{fig:B-inv}(a)), meaning the accumulation  of magnetic field at the reconnection site is the highest. Later on, in the two-plume case, we observe a flattening of the magnetic field around $x = 0$ mm, indicating a reduction of the magnetic flux accumulation.

We now quantitatively assess the reconnection rate. The production of magnetic field can be measured directly from the single plume case (Fig.\ref{fig:B-inv}(a)) with $\int B_y \D z = \SI{13}{T.mm} \sim B_0 \lambda$, where $B_0$ is the inflowing magnetic field and $\lambda$ is the extent (in the $z$ direction) of the plume that is reconnecting. Additionally, the magnetic flux is predominantly given by $\Phi = \int \!\!\! \int B_y \, \D x \D z$ and can numerically be calculated using Fig.\ref{fig:B-inv} at each time, yielding $\partial_t \Phi = 2.5 \pm \SI{0.6}{T.mm^2.ns^{-1}}$. Moreover, the velocity at which the front of the magnetic field expands is close to the Alfvén velocity (as confirmed by our FCI2 hydro-radiative simulation). This velocity is retrieved from the motion of the peak of the magnetic field for a single plume (as observed in Fig.\ref{fig:B-inv}(a-c)), yielding $V_0 = 400 \pm \SI{130}{km/s}$. Since the integrated form of the Maxwell-Faraday equation is $\partial_t \Phi = \lambda E$, we can also retrieve $E$, the convection electric field which drives the magnetic flux toward the current sheet.
It is used to obtain the dimensionless electric field $E^{\prime} = E/(B_0 V_0) = 0.48 \substack{+0.40 \\ -0.20}$. 
In the stationary case where the inflow of magnetic flux equals its outflow by reconnection, we have $E^{\prime} = E_r$ where $E_r$ is the reconnection rate.

This simple quantitative estimate indicates that the reconnection rate in such laser-driven reconnection exceeds commonly cited normalized reconnection rate of  fast reconnection, which is of the order of 0.1~\cite{cassak2017,Liu2017}. Such high rate may be the result induced by the dominant ram pressure, which forces the inflow of magnetic field into the MR region.

For the three-plume case, Fig.~\ref{fig:B-inv}(d) shows that, in stark contrast with the two-plume case, the local gradient of the path-integrated magnetic field stays much stronger. This is due to the magnetic strength at the outer edge of the CS being larger in three-plume case (see Fig.~\ref{fig:B-inv}(b-c)).
Since  the electron temperature and density are expected to be similar when driving a third plume, we can also assume that the rate of magnetic production will be similar in the two-plume and three-plume cases. Thus, we can therefore infer from Fig.~\ref{fig:B-inv}(d) that a stronger accumulation of the magnetic field occurs in the three-plume case with respect to the two-plume case.
This constitutes a direct evidence of the mitigation of MR in the case when multiple reconnection sites interact. Indeed, the magnetic field accumulates in the three-plume case because the annihilation induced by MR is then overall weakened. To quantitatively assess this weakening, we first measure the unperturbed magnetic flux. The latter can be measured on the opposite side of the MR site, i.e. it is the flux integrated between $x = \SI{0.9}{mm}$ and $\SI{2.6}{mm}$ in Fig.~\ref{fig:B-inv}(b-c)). Then, we can compare it to the quantity of reconnecting magnetic flux, which corresponds to the flux integrated between $\SI{0.0}{mm}$ and $\SI{0.9}{mm}$ in Fig.~\ref{fig:B-inv}(b-c). By comparing these two quantities, we estimate that, at 4.4 ns, 85\% of the flux has been annihilated~\footnote{In this configuration, the stationarity is not fulfilled, and thus the relation $E=E_r$ cannot be used.}. This is to be compared to this annihilation being close to 100\% in the two-plume case.

\begin{figure}[ht!]
\includegraphics[width=0.47\textwidth]{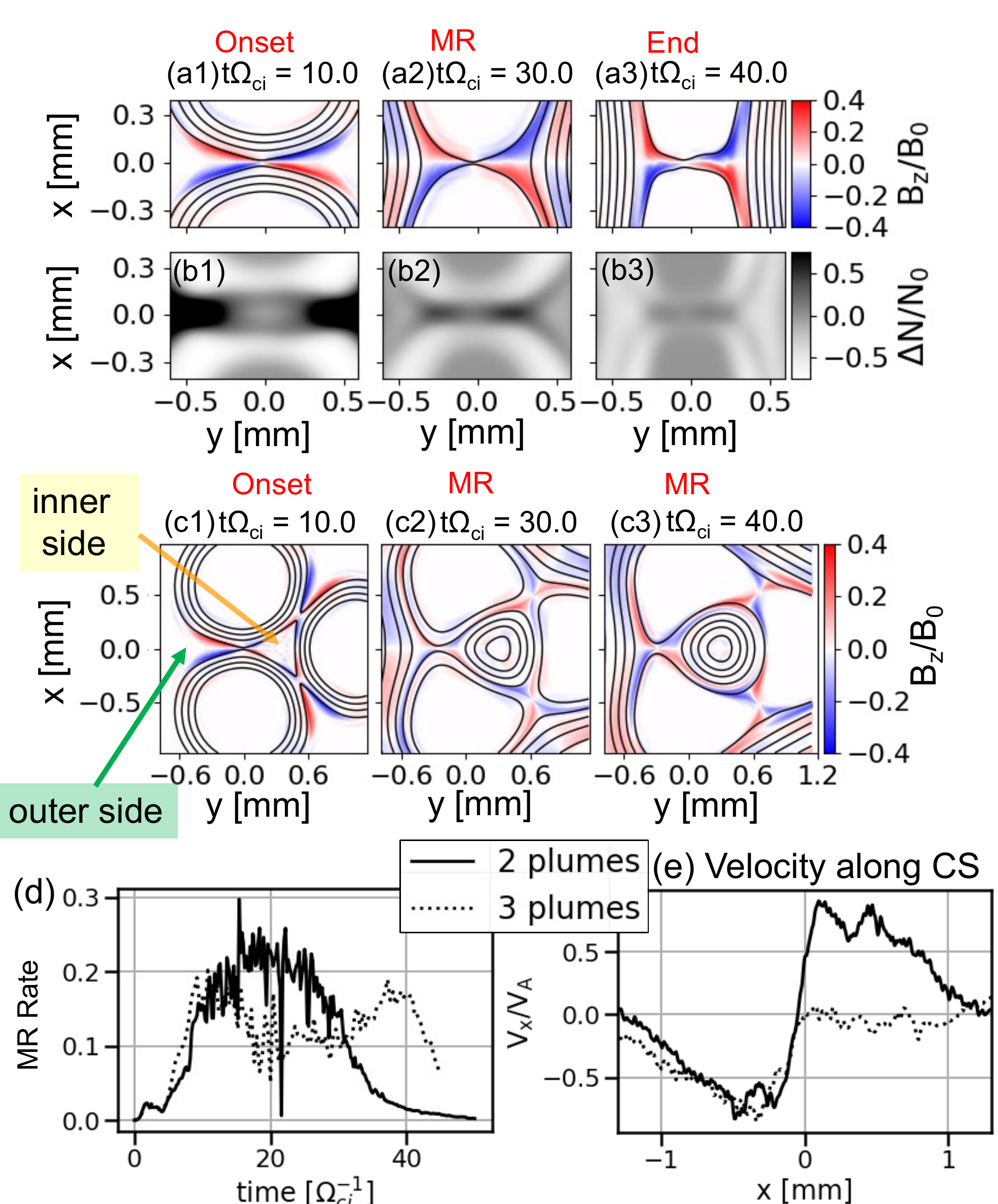}
\caption{\label{fig:simu} Results of hybrid-PIC simulations. Panels (a1-a3) display, at different times, the Hall magnetic field, for the two-plume case. Panels (b1-b3) display the corresponding synthetic proton radiography images (simulated by the ILZ code~\cite{Yao2022}). Panels (c1-c3) display the Hall magnetic field, for the three-plume case. The simulated temporal evolution of the reconnection rate is shown in panel (d) for the two (solid line) and three (dotted line) plasmas cases. Panel (e) displays the outflow velocity, computed at $t = \SI{25}{\Omega_{ci}^{-1}}$. }
\end{figure}

To unravel the underlying mechanism leading to the observed mitigation of  MR when it is at play simultaneously over adjacent sites, we performed two-dimensional hybrid PIC simulations using the {\sc heckle} code~\cite{heckle}.
Ions are kinetically treated as macro-particles (with a standard PIC method~\cite{Birdsall2018}), while the electrons are treated as a massless fluid. We use an isothermal closure equation.
It is analogous to maintaining the electron temperature using a laser pulse. 
The dissipation at the current sheet requires the introduction of kinetic ions for the onset of the fast reconnection ~\cite{birn2001+}. Two simulations, representative of the two investigated experimental cases, namely with two (Fig.\ref{fig:simu}(a-b)) and three (Fig.\ref{fig:simu}(c)) plasma plumes and their surrounding magnetic fields, are here detailed and compared. In the simulations, the magnetic field is initialized in a similar manner as in our previous numerical investigation~\cite{smets2014}, i.e. the size of the plumes and the strength of the magnetic field are obtained from the FCI2 simulations, which were benchmarked using the proton radiography images for the shot performed with a single plasma plume, i.e. when no MR is at play.

Fig.\ref{fig:simu}(a1-a3) shows snapshots of the $z$-component of the magnetic field ($B_z$) in the simulations for the two-plasmas case, from the onset to the end of MR. The corresponding synthetic proton images are shown in Fig.\ref{fig:simu}(b1-b3). At the MR onset,  magnetic compression is at its peak, i.e., the magnetic reversal is the steepest. This can be directly observed on the synthetic proton image, via the double line (mouth-like) pattern in Fig.\ref{fig:simu}(b1). Later on, the accumulated magnetic flux reconnects and gets evacuated, leading to a decrease of the contrast of the proton flux at the current sheet, similarly as observed in the experiment (see Fig.\ref{fig:rcf}(a1)). At the end of MR, i.e., when most of the magnetic flux has reconnected, the contrast of the probing proton is virtually null. This can be also observed in the experiment in the two-plume case at late time (see Fig.\ref{fig:rcf}(a2)).

The comparative evolution of $B_z$ and the magnetic flux in the simulations for the three-plume case is presented in Fig.\ref{fig:simu}(c1-c3). As for the two-plume simulation, we observe MR taking place. However, MR is here obviously slowed down, consistently with what is observed in the experiment: at late time (see Fig.\ref{fig:simu}(c3)), MR is still taking place as at earlier time. \SB{This slowing-down is also seen through a lower MR rate in the three-plume configuration, see Fig.\ref{fig:simu}(d).}

We attribute the delayed reconnection in the three-plume case to two mechanisms: (i) the hampered growth of the quadrupolar Hall field structure (depicted in color in Fig.\ref{fig:simu}(a1-a3) and (c1-c3)) that appears at the onset of MR~\cite{smets2014} and (ii) the accumulation of the trapped ejected plasma in between the three plumes, which  reduces the outflow velocity. The Hall field in the reconnection region is initiated as the ions get demagnetized while the electrons remain frozen in the plasma flow. This separation between the electron and ion trajectories induces a Hall current and, therefore, a quadrupolar magnetic field, which signature clearly appears at the onset of MR (see Fig.\ref{fig:simu}(a1,c1). We previously analyzed in Ref.~\cite{smets2014} that  such Hall structure is a prerequisite for MR to take place and concluded that its development  precedes MR. In this frame, we can understand the delayed reconnection in the three-plume case: it can be clearly seen in Fig.\ref{fig:simu}(c1) that the Hall field can develop normally in the outer zones (i.e. away from the central point of the overall assembly). However, it is disturbed in the inner side (i.e. in the triangular central structure) where there are opposite polarity Hall fields trying to grow at the same location. This prevents the Hall field to develop normally, as it does e.g. in Fig.\ref{fig:simu}(a1) and consequently destabilizes the unfolding of MR.

A second mechanism could contribute to the mitigation of MR   in the three-plume case: the trapping of the plasma downstream, in the inner side of the closed magnetic structure. As a consequence, half of the outflow at each reconnection site is trapped in the middle of the pattern, hence partially precluding the draining of the newly reconnected magnetic flux. This magnetic-trapped plasma hence acts like a cork, reducing the outflow velocity.  We measure in the simulations that the outflow velocity is about 
$\SI{0.75}{v_A}$, as shown in Fig.\ref{fig:simu}(e). However, in the inner side of the three-plume  structure, the outflow velocity is null. In the experiment we did not observe such a prominent closed magnetic structure. In contrast to the simulation, the experiment, being three-dimensional, is not infinitely invariant along the $z$-axis. Thus, the particles can escape and the “cork” effect is very likely not as important as in the two-dimensional simulation.

\begin{acknowledgments}
We thank the LMJ team for technical support, and Erik Lefebvre (CEA), Fabio Reale (INAF) and Alexandra Alexandrova (LPP) for discussions and comments. The PETAL laser was designed and constructed by CEA under the financial auspices of the Conseil Régional d’Aquitaine, the French Ministry of Research, and the European Union. The CRACC diagnostics were designed and commissioned on the LMJ-PETAL facility as a result of the PETAL+ project coordinated by Université de Bordeaux and funded by the French Agence Nationale de la Recherche under grant ANR-10-EQPX-42-01. This work was supported by the European Research Council (ERC) under the European Unions Horizon 2020 research and innovation program (Grant Agreement No. 787539). This work was partly done within the LABEX Plas@Par project and supported by Grant No. 11-IDEX- 0004-02 from ANR (France). The software used in this work was developed in part by the DOE NNSA-and DOE Office of Science-supported Flash Center for Computational Science at the University of Chicago and the University of Rochester. All data needed to evaluate the conclusions in the paper are present in the paper. Experimental data and simulations are respectively archived on servers at LULI and LPP laboratories and can be consulted upon request. 
\end{acknowledgments}

\bibliography{ref.bib}

\begin{thebibliography}{44}%
\makeatletter
\providecommand \@ifxundefined [1]{%
 \@ifx{#1\undefined}
}%
\providecommand \@ifnum [1]{%
 \ifnum #1\expandafter \@firstoftwo
 \else \expandafter \@secondoftwo
 \fi
}%
\providecommand \@ifx [1]{%
 \ifx #1\expandafter \@firstoftwo
 \else \expandafter \@secondoftwo
 \fi
}%
\providecommand \natexlab [1]{#1}%
\providecommand \enquote  [1]{``#1''}%
\providecommand \bibnamefont  [1]{#1}%
\providecommand \bibfnamefont [1]{#1}%
\providecommand \citenamefont [1]{#1}%
\providecommand \href@noop [0]{\@secondoftwo}%
\providecommand \href [0]{\begingroup \@sanitize@url \@href}%
\providecommand \@href[1]{\@@startlink{#1}\@@href}%
\providecommand \@@href[1]{\endgroup#1\@@endlink}%
\providecommand \@sanitize@url [0]{\catcode `\\12\catcode `\$12\catcode
  `\&12\catcode `\#12\catcode `\^12\catcode `\_12\catcode `\%12\relax}%
\providecommand \@@startlink[1]{}%
\providecommand \@@endlink[0]{}%
\providecommand \url  [0]{\begingroup\@sanitize@url \@url }%
\providecommand \@url [1]{\endgroup\@href {#1}{\urlprefix }}%
\providecommand \urlprefix  [0]{URL }%
\providecommand \Eprint [0]{\href }%
\providecommand \doibase [0]{https://doi.org/}%
\providecommand \selectlanguage [0]{\@gobble}%
\providecommand \bibinfo  [0]{\@secondoftwo}%
\providecommand \bibfield  [0]{\@secondoftwo}%
\providecommand \translation [1]{[#1]}%
\providecommand \BibitemOpen [0]{}%
\providecommand \bibitemStop [0]{}%
\providecommand \bibitemNoStop [0]{.\EOS\space}%
\providecommand \EOS [0]{\spacefactor3000\relax}%
\providecommand \BibitemShut  [1]{\csname bibitem#1\endcsname}%
\let\auto@bib@innerbib\@empty
\bibitem [{\citenamefont {{Parker}}(1957)}]{parker1957}%
  \BibitemOpen
  \bibfield  {author} {\bibinfo {author} {\bibfnamefont {E.~N.}\ \bibnamefont
  {{Parker}}},\ }\bibfield  {title} {\bibinfo {title} {Sweet's mechanism for
  merging magnetic fields in conducting fluids},\ }\href
  {https://doi.org/10.1029/JZ062i004p00509} {\bibfield  {journal} {\bibinfo
  {journal} {J. Geophys. Res.}\ }\textbf {\bibinfo {volume} {62}},\ \bibinfo
  {pages} {509} (\bibinfo {year} {1957})}\BibitemShut {NoStop}%
\bibitem [{\citenamefont {{Sweet}}(1958)}]{sweet1958}%
  \BibitemOpen
  \bibfield  {author} {\bibinfo {author} {\bibfnamefont {P.~A.}\ \bibnamefont
  {{Sweet}}},\ }\bibfield  {title} {\bibinfo {title} {The neutral point theory
  of solar flares},\ }in\ \href@noop {} {\emph {\bibinfo {booktitle}
  {Electromagnetic Phenomena in Cosmical Physics}}},\ \bibinfo {series} {IAU
  Symposium}, Vol.~\bibinfo {volume} {6},\ \bibinfo {editor} {edited by\
  \bibinfo {editor} {\bibfnamefont {B.}~\bibnamefont {{Lehnert}}}}\ (\bibinfo
  {year} {1958})\ p.\ \bibinfo {pages} {123}\BibitemShut {NoStop}%
\bibitem [{\citenamefont {{Petschek}}(1964)}]{petschek1964}%
  \BibitemOpen
  \bibfield  {author} {\bibinfo {author} {\bibfnamefont {H.~E.}\ \bibnamefont
  {{Petschek}}},\ }\bibfield  {title} {\bibinfo {title} {{Magnetic Field
  Annihilation}},\ }in\ \href@noop {} {\emph {\bibinfo {booktitle} {NASA
  Special Publication}}},\ Vol.~\bibinfo {volume} {50}\ (\bibinfo {year}
  {1964})\ p.\ \bibinfo {pages} {425}\BibitemShut {NoStop}%
\bibitem [{\citenamefont {Loureiro}\ and\ \citenamefont
  {Boldyrev}(2017)}]{Loureiro2017}%
  \BibitemOpen
  \bibfield  {author} {\bibinfo {author} {\bibfnamefont {N.~F.}\ \bibnamefont
  {Loureiro}}\ and\ \bibinfo {author} {\bibfnamefont {S.}~\bibnamefont
  {Boldyrev}},\ }\bibfield  {title} {\bibinfo {title} {Role of magnetic
  reconnection in magnetohydrodynamic turbulence},\ }\href
  {https://doi.org/10.1103/PhysRevLett.118.245101} {\bibfield  {journal}
  {\bibinfo  {journal} {\prl}\ }\textbf {\bibinfo {volume} {118}},\ \bibinfo
  {pages} {245101} (\bibinfo {year} {2017})}\BibitemShut {NoStop}%
\bibitem [{\citenamefont {Adhikari}\ \emph {et~al.}(2019)\citenamefont
  {Adhikari}, \citenamefont {Khabarova}, \citenamefont {Zank},\ and\
  \citenamefont {Zhao}}]{Adhikari2019}%
  \BibitemOpen
  \bibfield  {author} {\bibinfo {author} {\bibfnamefont {L.}~\bibnamefont
  {Adhikari}}, \bibinfo {author} {\bibfnamefont {O.}~\bibnamefont {Khabarova}},
  \bibinfo {author} {\bibfnamefont {G.~P.}\ \bibnamefont {Zank}},\ and\
  \bibinfo {author} {\bibfnamefont {L.-L.}\ \bibnamefont {Zhao}},\ }\bibfield
  {title} {\bibinfo {title} {The role of magnetic
  reconnection{\textendash}associated processes in local particle acceleration
  in the solar wind},\ }\href {https://doi.org/10.3847/1538-4357/ab05c6}
  {\bibfield  {journal} {\bibinfo  {journal} {The Astrophysical Journal}\
  }\textbf {\bibinfo {volume} {873}},\ \bibinfo {pages} {72} (\bibinfo {year}
  {2019})}\BibitemShut {NoStop}%
\bibitem [{\citenamefont {Drake}\ \emph {et~al.}(2010)\citenamefont {Drake},
  \citenamefont {Opher}, \citenamefont {Swisdak},\ and\ \citenamefont
  {Chamoun}}]{Drake2010}%
  \BibitemOpen
  \bibfield  {author} {\bibinfo {author} {\bibfnamefont {J.~F.}\ \bibnamefont
  {Drake}}, \bibinfo {author} {\bibfnamefont {M.}~\bibnamefont {Opher}},
  \bibinfo {author} {\bibfnamefont {M.}~\bibnamefont {Swisdak}},\ and\ \bibinfo
  {author} {\bibfnamefont {J.~N.}\ \bibnamefont {Chamoun}},\ }\bibfield
  {title} {\bibinfo {title} {A magnetic reconnection mechanism for the
  generation of anomalous cosmic rays},\ }\href
  {https://doi.org/10.1088/0004-637x/709/2/963} {\bibfield  {journal} {\bibinfo
   {journal} {The Astrophysical Journal}\ }\textbf {\bibinfo {volume} {709}},\
  \bibinfo {pages} {963} (\bibinfo {year} {2010})}\BibitemShut {NoStop}%
\bibitem [{\citenamefont {Guo}\ \emph {et~al.}(2015)\citenamefont {Guo},
  \citenamefont {Liu}, \citenamefont {Daughton},\ and\ \citenamefont
  {Li}}]{Guo2015}%
  \BibitemOpen
  \bibfield  {author} {\bibinfo {author} {\bibfnamefont {F.}~\bibnamefont
  {Guo}}, \bibinfo {author} {\bibfnamefont {Y.-H.}\ \bibnamefont {Liu}},
  \bibinfo {author} {\bibfnamefont {W.}~\bibnamefont {Daughton}},\ and\
  \bibinfo {author} {\bibfnamefont {H.}~\bibnamefont {Li}},\ }\bibfield
  {title} {\bibinfo {title} {Particle acceleration and plasma dynamics during
  magnetic reconnection in the magnetically dominated regime},\ }\href
  {https://doi.org/10.1088/0004-637x/806/2/167} {\bibfield  {journal} {\bibinfo
   {journal} {The Astrophysical Journal}\ }\textbf {\bibinfo {volume} {806}},\
  \bibinfo {pages} {167} (\bibinfo {year} {2015})}\BibitemShut {NoStop}%
\bibitem [{\citenamefont {Werner}\ and\ \citenamefont
  {Uzdensky}(2017)}]{Werner2017}%
  \BibitemOpen
  \bibfield  {author} {\bibinfo {author} {\bibfnamefont {G.~R.}\ \bibnamefont
  {Werner}}\ and\ \bibinfo {author} {\bibfnamefont {D.~A.}\ \bibnamefont
  {Uzdensky}},\ }\bibfield  {title} {\bibinfo {title} {Nonthermal particle
  acceleration in 3d relativistic magnetic reconnection in pair plasma},\
  }\href {https://doi.org/10.3847/2041-8213/aa7892} {\bibfield  {journal}
  {\bibinfo  {journal} {The Astrophysical Journal}\ }\textbf {\bibinfo {volume}
  {843}},\ \bibinfo {pages} {L27} (\bibinfo {year} {2017})}\BibitemShut
  {NoStop}%
\bibitem [{\citenamefont {{Vasyliunas}}(1975)}]{vasyliunas1975}%
  \BibitemOpen
  \bibfield  {author} {\bibinfo {author} {\bibfnamefont {V.~M.}\ \bibnamefont
  {{Vasyliunas}}},\ }\bibfield  {title} {\bibinfo {title} {Theoretical models
  of magnetic field line merging. i},\ }\href
  {https://doi.org/10.1029/RG013i001p00303} {\bibfield  {journal} {\bibinfo
  {journal} {Reviews of Geophysics and Space Physics}\ }\textbf {\bibinfo
  {volume} {13}},\ \bibinfo {pages} {303} (\bibinfo {year} {1975})}\BibitemShut
  {NoStop}%
\bibitem [{\citenamefont {{Cassak}}\ \emph {et~al.}(2017)\citenamefont
  {{Cassak}}, \citenamefont {{Liu}},\ and\ \citenamefont
  {{Shay}}}]{cassak2017}%
  \BibitemOpen
  \bibfield  {author} {\bibinfo {author} {\bibfnamefont {P.~A.}\ \bibnamefont
  {{Cassak}}}, \bibinfo {author} {\bibfnamefont {Y.~H.}\ \bibnamefont
  {{Liu}}},\ and\ \bibinfo {author} {\bibfnamefont {M.~A.}\ \bibnamefont
  {{Shay}}},\ }\bibfield  {title} {\bibinfo {title} {{A review of the 0.1
  reconnection rate problem}},\ }\href
  {https://doi.org/10.1017/S0022377817000666} {\bibfield  {journal} {\bibinfo
  {journal} {Journal of Plasma Physics}\ }\textbf {\bibinfo {volume} {83}},\
  \bibinfo {eid} {715830501} (\bibinfo {year} {2017})}\BibitemShut {NoStop}%
\bibitem [{\citenamefont {{Birn}}\ \emph {et~al.}(2001)\citenamefont {{Birn}}
  \emph {et~al.}}]{birn2001+}%
  \BibitemOpen
  \bibfield  {author} {\bibinfo {author} {\bibfnamefont {J.}~\bibnamefont
  {{Birn}}} \emph {et~al.},\ }\bibfield  {title} {\bibinfo {title} {Geospace
  environmental modeling (gem) magnetic reconnection challenge},\ }\href
  {https://doi.org/10.1029/1999JA900449} {\bibfield  {journal} {\bibinfo
  {journal} {J. Geophys. Res.}\ }\textbf {\bibinfo {volume} {106}},\ \bibinfo
  {pages} {3715} (\bibinfo {year} {2001})}\BibitemShut {NoStop}%
\bibitem [{\citenamefont {Ergun}\ \emph {et~al.}(2016)\citenamefont {Ergun}
  \emph {et~al.}}]{ergun2016}%
  \BibitemOpen
  \bibfield  {author} {\bibinfo {author} {\bibfnamefont {R.~E.}\ \bibnamefont
  {Ergun}} \emph {et~al.},\ }\bibfield  {title} {\bibinfo {title}
  {Magnetospheric multiscale satellites observations of parallel electric
  fields associated with magnetic reconnection},\ }\href
  {https://doi.org/10.1103/PhysRevLett.116.235102} {\bibfield  {journal}
  {\bibinfo  {journal} {\prl}\ }\textbf {\bibinfo {volume} {116}},\ \bibinfo
  {pages} {235102} (\bibinfo {year} {2016})}\BibitemShut {NoStop}%
\bibitem [{\citenamefont {Phan}\ \emph {et~al.}(2018)\citenamefont {Phan} \emph
  {et~al.}}]{phan2018}%
  \BibitemOpen
  \bibfield  {author} {\bibinfo {author} {\bibfnamefont {T.~D.}\ \bibnamefont
  {Phan}} \emph {et~al.},\ }\bibfield  {title} {\bibinfo {title} {Electron
  magnetic reconnection without ion coupling in earth's turbulent
  magnetosheath},\ }\href {https://doi.org/10.1038/s41586-018-0091-5}
  {\bibfield  {journal} {\bibinfo  {journal} {Nature}\ }\textbf {\bibinfo
  {volume} {557}},\ \bibinfo {pages} {202} (\bibinfo {year}
  {2018})}\BibitemShut {NoStop}%
\bibitem [{\citenamefont {Yamada}\ \emph {et~al.}(1997)\citenamefont {Yamada},
  \citenamefont {Ji}, \citenamefont {Hsu}, \citenamefont {Carter},
  \citenamefont {Kulsrud}, \citenamefont {Bretz}, \citenamefont {Jobes},
  \citenamefont {Ono},\ and\ \citenamefont {Perkins}}]{yamada1997}%
  \BibitemOpen
  \bibfield  {author} {\bibinfo {author} {\bibfnamefont {M.}~\bibnamefont
  {Yamada}}, \bibinfo {author} {\bibfnamefont {H.}~\bibnamefont {Ji}}, \bibinfo
  {author} {\bibfnamefont {S.}~\bibnamefont {Hsu}}, \bibinfo {author}
  {\bibfnamefont {T.}~\bibnamefont {Carter}}, \bibinfo {author} {\bibfnamefont
  {R.}~\bibnamefont {Kulsrud}}, \bibinfo {author} {\bibfnamefont
  {N.}~\bibnamefont {Bretz}}, \bibinfo {author} {\bibfnamefont
  {F.}~\bibnamefont {Jobes}}, \bibinfo {author} {\bibfnamefont
  {Y.}~\bibnamefont {Ono}},\ and\ \bibinfo {author} {\bibfnamefont
  {F.}~\bibnamefont {Perkins}},\ }\bibfield  {title} {\bibinfo {title} {Study
  of driven magnetic reconnection in a laboratory plasma},\ }\href
  {https://doi.org/10.1063/1.872336} {\bibfield  {journal} {\bibinfo  {journal}
  {\pop}\ }\textbf {\bibinfo {volume} {4}},\ \bibinfo {pages} {1936} (\bibinfo
  {year} {1997})}\BibitemShut {NoStop}%
\bibitem [{\citenamefont {{Egedal}}\ \emph {et~al.}(2007)\citenamefont
  {{Egedal}}, \citenamefont {{Fox}}, \citenamefont {{Katz}}, \citenamefont
  {{Porkolab}}, \citenamefont {{Reim}},\ and\ \citenamefont
  {{Zhang}}}]{egedal2007}%
  \BibitemOpen
  \bibfield  {author} {\bibinfo {author} {\bibfnamefont {J.}~\bibnamefont
  {{Egedal}}}, \bibinfo {author} {\bibfnamefont {W.}~\bibnamefont {{Fox}}},
  \bibinfo {author} {\bibfnamefont {N.}~\bibnamefont {{Katz}}}, \bibinfo
  {author} {\bibfnamefont {M.}~\bibnamefont {{Porkolab}}}, \bibinfo {author}
  {\bibfnamefont {K.}~\bibnamefont {{Reim}}},\ and\ \bibinfo {author}
  {\bibfnamefont {E.}~\bibnamefont {{Zhang}}},\ }\bibfield  {title} {\bibinfo
  {title} {Laboratory observations of spontaneous magnetic reconnection},\
  }\href {https://doi.org/10.1103/PhysRevLett.98.015003} {\bibfield  {journal}
  {\bibinfo  {journal} {Phys. Rev. Lett.}\ }\textbf {\bibinfo {volume} {98}},\
  \bibinfo {pages} {015003} (\bibinfo {year} {2007})}\BibitemShut {NoStop}%
\bibitem [{\citenamefont {Gekelman}\ \emph {et~al.}(2012)\citenamefont
  {Gekelman}, \citenamefont {Lawrence},\ and\ \citenamefont
  {Compernolle}}]{gekelman2012}%
  \BibitemOpen
  \bibfield  {author} {\bibinfo {author} {\bibfnamefont {W.}~\bibnamefont
  {Gekelman}}, \bibinfo {author} {\bibfnamefont {E.}~\bibnamefont {Lawrence}},\
  and\ \bibinfo {author} {\bibfnamefont {B.~V.}\ \bibnamefont {Compernolle}},\
  }\bibfield  {title} {\bibinfo {title} {Three-dimensional reconnection
  involving magnetic flux ropes},\ }\href
  {https://doi.org/10.1088/0004-637x/753/2/131} {\bibfield  {journal} {\bibinfo
   {journal} {The Astrophysical Journal}\ }\textbf {\bibinfo {volume} {753}},\
  \bibinfo {pages} {131} (\bibinfo {year} {2012})}\BibitemShut {NoStop}%
\bibitem [{\citenamefont {Fox}\ \emph {et~al.}(2018)\citenamefont {Fox},
  \citenamefont {Wilder}, \citenamefont {Eriksson}, \citenamefont
  {Jara-Almonte}, \citenamefont {Pucci}, \citenamefont {Yoo}, \citenamefont
  {Ji}, \citenamefont {Yamada}, \citenamefont {Ergun}, \citenamefont
  {Oieroset},\ and\ \citenamefont {Phan}}]{fox2018}%
  \BibitemOpen
  \bibfield  {author} {\bibinfo {author} {\bibfnamefont {W.}~\bibnamefont
  {Fox}}, \bibinfo {author} {\bibfnamefont {F.~D.}\ \bibnamefont {Wilder}},
  \bibinfo {author} {\bibfnamefont {S.}~\bibnamefont {Eriksson}}, \bibinfo
  {author} {\bibfnamefont {J.}~\bibnamefont {Jara-Almonte}}, \bibinfo {author}
  {\bibfnamefont {F.}~\bibnamefont {Pucci}}, \bibinfo {author} {\bibfnamefont
  {J.}~\bibnamefont {Yoo}}, \bibinfo {author} {\bibfnamefont {H.}~\bibnamefont
  {Ji}}, \bibinfo {author} {\bibfnamefont {M.}~\bibnamefont {Yamada}}, \bibinfo
  {author} {\bibfnamefont {R.~E.}\ \bibnamefont {Ergun}}, \bibinfo {author}
  {\bibfnamefont {M.}~\bibnamefont {Oieroset}},\ and\ \bibinfo {author}
  {\bibfnamefont {T.~D.}\ \bibnamefont {Phan}},\ }\bibfield  {title} {\bibinfo
  {title} {Energy conversion by parallel electric fields during guide field
  reconnection in scaled laboratory and space experiments},\ }\href
  {https://doi.org/10.1029/2018gl079883} {\bibfield  {journal} {\bibinfo
  {journal} {Geophysical Research Letters}\ }\textbf {\bibinfo {volume} {45}},\
  \bibinfo {pages} {12677} (\bibinfo {year} {2018})}\BibitemShut {NoStop}%
\bibitem [{\citenamefont {{Fox}}\ \emph {et~al.}(2011)\citenamefont {{Fox}},
  \citenamefont {{Bhattacharjee}},\ and\ \citenamefont
  {{Germaschewski}}}]{fox2011}%
  \BibitemOpen
  \bibfield  {author} {\bibinfo {author} {\bibfnamefont {W.}~\bibnamefont
  {{Fox}}}, \bibinfo {author} {\bibfnamefont {A.}~\bibnamefont
  {{Bhattacharjee}}},\ and\ \bibinfo {author} {\bibfnamefont {K.}~\bibnamefont
  {{Germaschewski}}},\ }\bibfield  {title} {\bibinfo {title} {Fast magnetic
  reconnection in laser-produced plasma bubbles},\ }\href
  {https://doi.org/10.1103/PhysRevLett.106.215003} {\bibfield  {journal}
  {\bibinfo  {journal} {Phys. Rev. Letter}\ }\textbf {\bibinfo {volume}
  {106}},\ \bibinfo {pages} {215003} (\bibinfo {year} {2011})}\BibitemShut
  {NoStop}%
\bibitem [{\citenamefont {{Smets}}\ \emph {et~al.}(2014)\citenamefont
  {{Smets}}, \citenamefont {{Aunai}}, \citenamefont {{Belmont}}, \citenamefont
  {{Boniface}},\ and\ \citenamefont {{Fuchs}}}]{smets2014}%
  \BibitemOpen
  \bibfield  {author} {\bibinfo {author} {\bibfnamefont {R.}~\bibnamefont
  {{Smets}}}, \bibinfo {author} {\bibfnamefont {N.}~\bibnamefont {{Aunai}}},
  \bibinfo {author} {\bibfnamefont {G.}~\bibnamefont {{Belmont}}}, \bibinfo
  {author} {\bibfnamefont {C.}~\bibnamefont {{Boniface}}},\ and\ \bibinfo
  {author} {\bibfnamefont {J.}~\bibnamefont {{Fuchs}}},\ }\bibfield  {title}
  {\bibinfo {title} {On the relationship between quadrupolar magnetic field and
  collisionless reconnection},\ }\href {https://doi.org/10.1063/1.4885097}
  {\bibfield  {journal} {\bibinfo  {journal} {Phys. Plasmas}\ }\textbf
  {\bibinfo {volume} {21}},\ \bibinfo {pages} {062111} (\bibinfo {year}
  {2014})},\ \bibinfo {note} {none}\BibitemShut {NoStop}%
\bibitem [{\citenamefont {{Nilson}}\ \emph {et~al.}(2006)\citenamefont
  {{Nilson}} \emph {et~al.}}]{nilson2006}%
  \BibitemOpen
  \bibfield  {author} {\bibinfo {author} {\bibfnamefont {P.~M.}\ \bibnamefont
  {{Nilson}}} \emph {et~al.},\ }\bibfield  {title} {\bibinfo {title} {Magnetic
  reconnection and plasma dynamics in two-beam laser-solid interactions},\
  }\href {https://doi.org/10.1103/PhysRevLett.97.255001} {\bibfield  {journal}
  {\bibinfo  {journal} {Phys. Rev. Lett.}\ }\textbf {\bibinfo {volume} {97}},\
  \bibinfo {pages} {255001} (\bibinfo {year} {2006})}\BibitemShut {NoStop}%
\bibitem [{\citenamefont {Li}\ \emph {et~al.}(2007)\citenamefont {Li},
  \citenamefont {S{\'{e}}guin}, \citenamefont {Frenje}, \citenamefont {Rygg},
  \citenamefont {Petrasso}, \citenamefont {Town}, \citenamefont {Landen},
  \citenamefont {Knauer},\ and\ \citenamefont {Smalyuk}}]{Li2007}%
  \BibitemOpen
  \bibfield  {author} {\bibinfo {author} {\bibfnamefont {C.~K.}\ \bibnamefont
  {Li}}, \bibinfo {author} {\bibfnamefont {F.~H.}\ \bibnamefont
  {S{\'{e}}guin}}, \bibinfo {author} {\bibfnamefont {J.~A.}\ \bibnamefont
  {Frenje}}, \bibinfo {author} {\bibfnamefont {J.~R.}\ \bibnamefont {Rygg}},
  \bibinfo {author} {\bibfnamefont {R.~D.}\ \bibnamefont {Petrasso}}, \bibinfo
  {author} {\bibfnamefont {R.~P.~J.}\ \bibnamefont {Town}}, \bibinfo {author}
  {\bibfnamefont {O.~L.}\ \bibnamefont {Landen}}, \bibinfo {author}
  {\bibfnamefont {J.~P.}\ \bibnamefont {Knauer}},\ and\ \bibinfo {author}
  {\bibfnamefont {V.~A.}\ \bibnamefont {Smalyuk}},\ }\bibfield  {title}
  {\bibinfo {title} {Observation of megagauss-field topology changes due to
  magnetic reconnection in laser-produced plasmas},\ }\href
  {https://doi.org/10.1103/physrevlett.99.055001} {\bibfield  {journal}
  {\bibinfo  {journal} {\prl}\ }\textbf {\bibinfo {volume} {99}},\ \bibinfo
  {pages} {055001} (\bibinfo {year} {2007})}\BibitemShut {NoStop}%
\bibitem [{\citenamefont {{Loureiro}}\ \emph {et~al.}(2012)\citenamefont
  {{Loureiro}}, \citenamefont {{Samtaney}}, \citenamefont {{Schekochihin}},\
  and\ \citenamefont {{Uzdensky}}}]{loureiro2012}%
  \BibitemOpen
  \bibfield  {author} {\bibinfo {author} {\bibfnamefont {N.~F.}\ \bibnamefont
  {{Loureiro}}}, \bibinfo {author} {\bibfnamefont {R.}~\bibnamefont
  {{Samtaney}}}, \bibinfo {author} {\bibfnamefont {A.~A.}\ \bibnamefont
  {{Schekochihin}}},\ and\ \bibinfo {author} {\bibfnamefont {D.~A.}\
  \bibnamefont {{Uzdensky}}},\ }\bibfield  {title} {\bibinfo {title} {Magnetic
  reconnection and stochastic plasmoid chains in high-lundquist-number
  plasmas},\ }\href {https://doi.org/10.1063/1.3703318} {\bibfield  {journal}
  {\bibinfo  {journal} {\pop}\ }\textbf {\bibinfo {volume} {19}},\ \bibinfo
  {pages} {042303} (\bibinfo {year} {2012})}\BibitemShut {NoStop}%
\bibitem [{\citenamefont {Dattolo}\ \emph {et~al.}(2001)\citenamefont
  {Dattolo}, \citenamefont {Suter}, \citenamefont {Monteil}, \citenamefont
  {Jadaud}, \citenamefont {Dague}, \citenamefont {Glenzer}, \citenamefont
  {Turner}, \citenamefont {Juraszek}, \citenamefont {Lasinski}, \citenamefont
  {Decker}, \citenamefont {Landen},\ and\ \citenamefont
  {MacGowan}}]{Dattolo2001}%
  \BibitemOpen
  \bibfield  {author} {\bibinfo {author} {\bibfnamefont {E.}~\bibnamefont
  {Dattolo}}, \bibinfo {author} {\bibfnamefont {L.}~\bibnamefont {Suter}},
  \bibinfo {author} {\bibfnamefont {M.~C.}\ \bibnamefont {Monteil}}, \bibinfo
  {author} {\bibfnamefont {J.~P.}\ \bibnamefont {Jadaud}}, \bibinfo {author}
  {\bibfnamefont {N.}~\bibnamefont {Dague}}, \bibinfo {author} {\bibfnamefont
  {S.}~\bibnamefont {Glenzer}}, \bibinfo {author} {\bibfnamefont
  {R.}~\bibnamefont {Turner}}, \bibinfo {author} {\bibfnamefont
  {D.}~\bibnamefont {Juraszek}}, \bibinfo {author} {\bibfnamefont
  {B.}~\bibnamefont {Lasinski}}, \bibinfo {author} {\bibfnamefont
  {C.}~\bibnamefont {Decker}}, \bibinfo {author} {\bibfnamefont
  {O.}~\bibnamefont {Landen}},\ and\ \bibinfo {author} {\bibfnamefont
  {B.}~\bibnamefont {MacGowan}},\ }\bibfield  {title} {\bibinfo {title} {Status
  of our understanding and modeling of x-ray coupling efficiency in laser
  heated hohlraums},\ }\href {https://doi.org/10.1063/1.1324659} {\bibfield
  {journal} {\bibinfo  {journal} {\pop}\ }\textbf {\bibinfo {volume} {8}},\
  \bibinfo {pages} {260} (\bibinfo {year} {2001})}\BibitemShut {NoStop}%
\bibitem [{\citenamefont {Schlüter}(1950)}]{Biermann1950}%
  \BibitemOpen
  \bibfield  {author} {\bibinfo {author} {\bibfnamefont {A.}~\bibnamefont
  {Schlüter}},\ }\bibfield  {title} {\bibinfo {title} {Über den ursprung der
  magnetfelder auf sternen und im interstellaren raum},\ }\href
  {https://doi.org/doi:10.1515/zna-1950-0201} {\bibfield  {journal} {\bibinfo
  {journal} {Zeitschrift für Naturforschung A}\ }\textbf {\bibinfo {volume}
  {5}},\ \bibinfo {pages} {65} (\bibinfo {year} {1950})}\BibitemShut {NoStop}%
\bibitem [{\citenamefont {Matteucci}\ \emph {et~al.}(2018)\citenamefont
  {Matteucci}, \citenamefont {Fox}, \citenamefont {Bhattacharjee},
  \citenamefont {Schaeffer}, \citenamefont {Moissard}, \citenamefont
  {Germaschewski}, \citenamefont {Fiksel},\ and\ \citenamefont
  {Hu}}]{Matteucci2018}%
  \BibitemOpen
  \bibfield  {author} {\bibinfo {author} {\bibfnamefont {J.}~\bibnamefont
  {Matteucci}}, \bibinfo {author} {\bibfnamefont {W.}~\bibnamefont {Fox}},
  \bibinfo {author} {\bibfnamefont {A.}~\bibnamefont {Bhattacharjee}}, \bibinfo
  {author} {\bibfnamefont {D.~B.}\ \bibnamefont {Schaeffer}}, \bibinfo {author}
  {\bibfnamefont {C.}~\bibnamefont {Moissard}}, \bibinfo {author}
  {\bibfnamefont {K.}~\bibnamefont {Germaschewski}}, \bibinfo {author}
  {\bibfnamefont {G.}~\bibnamefont {Fiksel}},\ and\ \bibinfo {author}
  {\bibfnamefont {S.~X.}\ \bibnamefont {Hu}},\ }\bibfield  {title} {\bibinfo
  {title} {Biermann-battery-mediated magnetic reconnection in 3d colliding
  plasmas},\ }\href {https://doi.org/10.1103/PhysRevLett.121.095001} {\bibfield
   {journal} {\bibinfo  {journal} {\prl}\ }\textbf {\bibinfo {volume} {121}},\
  \bibinfo {pages} {095001} (\bibinfo {year} {2018})}\BibitemShut {NoStop}%
\bibitem [{\citenamefont {Burlaga}\ \emph {et~al.}(2002)\citenamefont
  {Burlaga}, \citenamefont {Plunkett},\ and\ \citenamefont
  {St.~Cyr}}]{Burlaga2002}%
  \BibitemOpen
  \bibfield  {author} {\bibinfo {author} {\bibfnamefont {L.~F.}\ \bibnamefont
  {Burlaga}}, \bibinfo {author} {\bibfnamefont {S.~P.}\ \bibnamefont
  {Plunkett}},\ and\ \bibinfo {author} {\bibfnamefont {O.~C.}\ \bibnamefont
  {St.~Cyr}},\ }\bibfield  {title} {\bibinfo {title} {Successive cmes and
  complex ejecta},\ }\href {https://doi.org/10.1029/2001JA000255} {\bibfield
  {journal} {\bibinfo  {journal} {Journal of Geophysical Research (Space
  Physics)}\ }\textbf {\bibinfo {volume} {107}},\ \bibinfo {eid} {1266}
  (\bibinfo {year} {2002})}\BibitemShut {NoStop}%
\bibitem [{\citenamefont {{Bransgrove}}\ \emph {et~al.}(2021)\citenamefont
  {{Bransgrove}}, \citenamefont {{Ripperda}},\ and\ \citenamefont
  {{Philippov}}}]{Bransgrove2021}%
  \BibitemOpen
  \bibfield  {author} {\bibinfo {author} {\bibfnamefont {A.}~\bibnamefont
  {{Bransgrove}}}, \bibinfo {author} {\bibfnamefont {B.}~\bibnamefont
  {{Ripperda}}},\ and\ \bibinfo {author} {\bibfnamefont {A.}~\bibnamefont
  {{Philippov}}},\ }\bibfield  {title} {\bibinfo {title} {{Magnetic Hair and
  Reconnection in Black Hole Magnetospheres}},\ }\href
  {https://doi.org/10.1103/PhysRevLett.127.055101} {\bibfield  {journal}
  {\bibinfo  {journal} {\prl}\ }\textbf {\bibinfo {volume} {127}},\ \bibinfo
  {eid} {055101} (\bibinfo {year} {2021})}\BibitemShut {NoStop}%
\bibitem [{\citenamefont {Casner}\ \emph {et~al.}(2015)\citenamefont {Casner},
  \citenamefont {Caillaud}, \citenamefont {Darbon}, \citenamefont {Duval},
  \citenamefont {Thfouin}, \citenamefont {Jadaud}, \citenamefont {LeBreton},
  \citenamefont {Reverdin}, \citenamefont {Rosse}, \citenamefont {Rosch},
  \citenamefont {Blanchot}, \citenamefont {Villette}, \citenamefont {Wrobel},\
  and\ \citenamefont {Miquel}}]{Casner2015}%
  \BibitemOpen
  \bibfield  {author} {\bibinfo {author} {\bibfnamefont {A.}~\bibnamefont
  {Casner}}, \bibinfo {author} {\bibfnamefont {T.}~\bibnamefont {Caillaud}},
  \bibinfo {author} {\bibfnamefont {S.}~\bibnamefont {Darbon}}, \bibinfo
  {author} {\bibfnamefont {A.}~\bibnamefont {Duval}}, \bibinfo {author}
  {\bibfnamefont {I.}~\bibnamefont {Thfouin}}, \bibinfo {author} {\bibfnamefont
  {J.}~\bibnamefont {Jadaud}}, \bibinfo {author} {\bibfnamefont
  {J.}~\bibnamefont {LeBreton}}, \bibinfo {author} {\bibfnamefont
  {C.}~\bibnamefont {Reverdin}}, \bibinfo {author} {\bibfnamefont
  {B.}~\bibnamefont {Rosse}}, \bibinfo {author} {\bibfnamefont
  {R.}~\bibnamefont {Rosch}}, \bibinfo {author} {\bibfnamefont
  {N.}~\bibnamefont {Blanchot}}, \bibinfo {author} {\bibfnamefont
  {B.}~\bibnamefont {Villette}}, \bibinfo {author} {\bibfnamefont
  {R.}~\bibnamefont {Wrobel}},\ and\ \bibinfo {author} {\bibfnamefont
  {J.}~\bibnamefont {Miquel}},\ }\bibfield  {title} {\bibinfo {title}
  {{LMJ}/{PETAL} laser facility: Overview and opportunities for laboratory
  astrophysics},\ }\href {https://doi.org/10.1016/j.hedp.2014.11.009}
  {\bibfield  {journal} {\bibinfo  {journal} {High Energy Density Physics}\
  }\textbf {\bibinfo {volume} {17}},\ \bibinfo {pages} {2} (\bibinfo {year}
  {2015})}\BibitemShut {NoStop}%
\bibitem [{\citenamefont {{Denis}}\ \emph {et~al.}(2021)\citenamefont
  {{Denis}}, \citenamefont {{Nicolaizeau}}, \citenamefont {{N{\'e}auport}},
  \citenamefont {{Lacombe}},\ and\ \citenamefont {{Fourtillan}}}]{Denis2021}%
  \BibitemOpen
  \bibfield  {author} {\bibinfo {author} {\bibfnamefont {V.}~\bibnamefont
  {{Denis}}}, \bibinfo {author} {\bibfnamefont {M.}~\bibnamefont
  {{Nicolaizeau}}}, \bibinfo {author} {\bibfnamefont {J.}~\bibnamefont
  {{N{\'e}auport}}}, \bibinfo {author} {\bibfnamefont {C.}~\bibnamefont
  {{Lacombe}}},\ and\ \bibinfo {author} {\bibfnamefont {P.}~\bibnamefont
  {{Fourtillan}}},\ }\bibfield  {title} {\bibinfo {title} {{LMJ 2021 facility
  status}},\ }in\ \href {https://doi.org/10.1117/12.2576671} {\emph {\bibinfo
  {booktitle} {High Power Lasers for Fusion Research VI}}},\ \bibinfo {series}
  {Society of Photo-Optical Instrumentation Engineers (SPIE) Conference
  Series}, Vol.\ \bibinfo {volume} {11666},\ \bibinfo {editor} {edited by\
  \bibinfo {editor} {\bibfnamefont {A.~A.~S.}\ \bibnamefont {{Awwal}}}\ and\
  \bibinfo {editor} {\bibfnamefont {C.~L.}\ \bibnamefont {{Haefner}}}}\
  (\bibinfo {year} {2021})\ p.\ \bibinfo {pages} {1166603}\BibitemShut
  {NoStop}%
\bibitem [{\citenamefont {Schaeffer}\ \emph {et~al.}(2022)\citenamefont
  {Schaeffer} \emph {et~al.}}]{reviewSchaeffer}%
  \BibitemOpen
  \bibfield  {author} {\bibinfo {author} {\bibnamefont {Schaeffer}} \emph
  {et~al.},\ }\href {https://doi.org/10.48550/ARXIV.2212.08252} {\bibinfo
  {title} {Proton imaging of high-energy-density laboratory plasmas}} (\bibinfo
  {year} {2022})\BibitemShut {NoStop}%
\bibitem [{\citenamefont {Wilks}\ \emph {et~al.}(2001)\citenamefont {Wilks},
  \citenamefont {Langdon}, \citenamefont {Cowan}, \citenamefont {Roth},
  \citenamefont {Singh}, \citenamefont {Hatchett}, \citenamefont {Key},
  \citenamefont {Pennington}, \citenamefont {MacKinnon},\ and\ \citenamefont
  {Snavely}}]{Wilks2001}%
  \BibitemOpen
  \bibfield  {author} {\bibinfo {author} {\bibfnamefont {S.~C.}\ \bibnamefont
  {Wilks}}, \bibinfo {author} {\bibfnamefont {A.~B.}\ \bibnamefont {Langdon}},
  \bibinfo {author} {\bibfnamefont {T.~E.}\ \bibnamefont {Cowan}}, \bibinfo
  {author} {\bibfnamefont {M.}~\bibnamefont {Roth}}, \bibinfo {author}
  {\bibfnamefont {M.}~\bibnamefont {Singh}}, \bibinfo {author} {\bibfnamefont
  {S.}~\bibnamefont {Hatchett}}, \bibinfo {author} {\bibfnamefont {M.~H.}\
  \bibnamefont {Key}}, \bibinfo {author} {\bibfnamefont {D.}~\bibnamefont
  {Pennington}}, \bibinfo {author} {\bibfnamefont {A.}~\bibnamefont
  {MacKinnon}},\ and\ \bibinfo {author} {\bibfnamefont {R.~A.}\ \bibnamefont
  {Snavely}},\ }\bibfield  {title} {\bibinfo {title} {Energetic proton
  generation in ultra-intense laser{\textendash}solid interactions},\ }\href
  {https://doi.org/10.1063/1.1333697} {\bibfield  {journal} {\bibinfo
  {journal} {\pop}\ }\textbf {\bibinfo {volume} {8}},\ \bibinfo {pages} {542}
  (\bibinfo {year} {2001})}\BibitemShut {NoStop}%
\bibitem [{\citenamefont {Raffestin}\ \emph {et~al.}(2021)\citenamefont
  {Raffestin} \emph {et~al.}}]{Raffestin2021}%
  \BibitemOpen
  \bibfield  {author} {\bibinfo {author} {\bibfnamefont {D.}~\bibnamefont
  {Raffestin}} \emph {et~al.},\ }\bibfield  {title} {\bibinfo {title} {Enhanced
  ion acceleration using the high-energy petawatt {PETAL} laser},\ }\href
  {https://doi.org/10.1063/5.0046679} {\bibfield  {journal} {\bibinfo
  {journal} {Matter and Radiation at Extremes}\ }\textbf {\bibinfo {volume}
  {6}},\ \bibinfo {pages} {056901} (\bibinfo {year} {2021})}\BibitemShut
  {NoStop}%
\bibitem [{\citenamefont {Cecchetti}\ \emph {et~al.}(2009)\citenamefont
  {Cecchetti} \emph {et~al.}}]{Cecchetti2009}%
  \BibitemOpen
  \bibfield  {author} {\bibinfo {author} {\bibfnamefont {C.~A.}\ \bibnamefont
  {Cecchetti}} \emph {et~al.},\ }\bibfield  {title} {\bibinfo {title} {Magnetic
  field measurements in laser-produced plasmas via proton deflectometry},\
  }\href {https://doi.org/10.1063/1.3097899} {\bibfield  {journal} {\bibinfo
  {journal} {\pop}\ }\textbf {\bibinfo {volume} {16}},\ \bibinfo {pages}
  {043102} (\bibinfo {year} {2009})}\BibitemShut {NoStop}%
\bibitem [{\citenamefont {Petrasso}\ \emph {et~al.}(2009)\citenamefont
  {Petrasso} \emph {et~al.}}]{Petrasso2009}%
  \BibitemOpen
  \bibfield  {author} {\bibinfo {author} {\bibfnamefont {R.~D.}\ \bibnamefont
  {Petrasso}} \emph {et~al.},\ }\bibfield  {title} {\bibinfo {title} {Lorentz
  mapping of magnetic fields in hot dense plasmas},\ }\href
  {https://doi.org/10.1103/physrevlett.103.085001} {\bibfield  {journal}
  {\bibinfo  {journal} {\prl}\ }\textbf {\bibinfo {volume} {103}},\ \bibinfo
  {pages} {085001} (\bibinfo {year} {2009})}\BibitemShut {NoStop}%
\bibitem [{\citenamefont {Lancia}\ \emph {et~al.}(2014)\citenamefont {Lancia}
  \emph {et~al.}}]{Lancia2014}%
  \BibitemOpen
  \bibfield  {author} {\bibinfo {author} {\bibfnamefont {L.}~\bibnamefont
  {Lancia}} \emph {et~al.},\ }\bibfield  {title} {\bibinfo {title} {Topology of
  megagauss magnetic fields and of heat-carrying electrons produced in a
  high-power laser-solid interaction},\ }\href
  {https://doi.org/10.1103/physrevlett.113.235001} {\bibfield  {journal}
  {\bibinfo  {journal} {\prl}\ }\textbf {\bibinfo {volume} {113}},\ \bibinfo
  {pages} {235001} (\bibinfo {year} {2014})}\BibitemShut {NoStop}%
\bibitem [{\citenamefont {Campbell}\ \emph {et~al.}(2020)\citenamefont
  {Campbell}, \citenamefont {Walsh}, \citenamefont {Russell}, \citenamefont
  {Chittenden}, \citenamefont {Crilly}, \citenamefont {Fiksel}, \citenamefont
  {Nilson}, \citenamefont {Thomas}, \citenamefont {Krushelnick},\ and\
  \citenamefont {Willingale}}]{Campbell2020}%
  \BibitemOpen
  \bibfield  {author} {\bibinfo {author} {\bibfnamefont {P.}~\bibnamefont
  {Campbell}}, \bibinfo {author} {\bibfnamefont {C.}~\bibnamefont {Walsh}},
  \bibinfo {author} {\bibfnamefont {B.}~\bibnamefont {Russell}}, \bibinfo
  {author} {\bibfnamefont {J.}~\bibnamefont {Chittenden}}, \bibinfo {author}
  {\bibfnamefont {A.}~\bibnamefont {Crilly}}, \bibinfo {author} {\bibfnamefont
  {G.}~\bibnamefont {Fiksel}}, \bibinfo {author} {\bibfnamefont
  {P.}~\bibnamefont {Nilson}}, \bibinfo {author} {\bibfnamefont
  {A.}~\bibnamefont {Thomas}}, \bibinfo {author} {\bibfnamefont
  {K.}~\bibnamefont {Krushelnick}},\ and\ \bibinfo {author} {\bibfnamefont
  {L.}~\bibnamefont {Willingale}},\ }\bibfield  {title} {\bibinfo {title}
  {Magnetic signatures of radiation-driven double ablation fronts},\ }\href
  {https://doi.org/10.1103/physrevlett.125.145001} {\bibfield  {journal}
  {\bibinfo  {journal} {\prl}\ }\textbf {\bibinfo {volume} {125}},\ \bibinfo
  {pages} {145001} (\bibinfo {year} {2020})}\BibitemShut {NoStop}%
\bibitem [{\citenamefont {{Bola{\~n}os}}\ \emph {et~al.}(2022)\citenamefont
  {{Bola{\~n}os}} \emph {et~al.}}]{Bolanos2022}%
  \BibitemOpen
  \bibfield  {author} {\bibinfo {author} {\bibfnamefont {S.}~\bibnamefont
  {{Bola{\~n}os}}} \emph {et~al.},\ }\bibfield  {title} {\bibinfo {title}
  {{Laboratory evidence of magnetic reconnection hampered in obliquely
  interacting flux tubes}},\ }\href
  {https://doi.org/10.1038/s41467-022-33813-9} {\bibfield  {journal} {\bibinfo
  {journal} {Nature Communications}\ }\textbf {\bibinfo {volume} {13}},\
  \bibinfo {eid} {6426} (\bibinfo {year} {2022})}\BibitemShut {NoStop}%
\bibitem [{\citenamefont {Bott}\ \emph {et~al.}(2017)\citenamefont {Bott},
  \citenamefont {Graziani}, \citenamefont {Tzeferacos}, \citenamefont {White},
  \citenamefont {Lamb}, \citenamefont {Gregori},\ and\ \citenamefont
  {Schekochihin}}]{Bott2017}%
  \BibitemOpen
  \bibfield  {author} {\bibinfo {author} {\bibfnamefont {A.~F.~A.}\
  \bibnamefont {Bott}}, \bibinfo {author} {\bibfnamefont {C.}~\bibnamefont
  {Graziani}}, \bibinfo {author} {\bibfnamefont {P.}~\bibnamefont
  {Tzeferacos}}, \bibinfo {author} {\bibfnamefont {T.~G.}\ \bibnamefont
  {White}}, \bibinfo {author} {\bibfnamefont {D.~Q.}\ \bibnamefont {Lamb}},
  \bibinfo {author} {\bibfnamefont {G.}~\bibnamefont {Gregori}},\ and\ \bibinfo
  {author} {\bibfnamefont {A.~A.}\ \bibnamefont {Schekochihin}},\ }\bibfield
  {title} {\bibinfo {title} {Proton imaging of stochastic magnetic fields},\
  }\href {https://doi.org/10.1017/s0022377817000939} {\bibfield  {journal}
  {\bibinfo  {journal} {Journal of Plasma Physics}\ }\textbf {\bibinfo {volume}
  {83}},\ \bibinfo {pages} {905830614} (\bibinfo {year} {2017})}\BibitemShut
  {NoStop}%
\bibitem [{\citenamefont {Graziani}\ \emph {et~al.}(2017)\citenamefont
  {Graziani}, \citenamefont {Tzeferacos}, \citenamefont {Lamb},\ and\
  \citenamefont {Li}}]{Graziani2017}%
  \BibitemOpen
  \bibfield  {author} {\bibinfo {author} {\bibfnamefont {C.}~\bibnamefont
  {Graziani}}, \bibinfo {author} {\bibfnamefont {P.}~\bibnamefont
  {Tzeferacos}}, \bibinfo {author} {\bibfnamefont {D.~Q.}\ \bibnamefont
  {Lamb}},\ and\ \bibinfo {author} {\bibfnamefont {C.}~\bibnamefont {Li}},\
  }\bibfield  {title} {\bibinfo {title} {Inferring morphology and strength of
  magnetic fields from proton radiographs},\ }\href
  {https://doi.org/10.1063/1.5013029} {\bibfield  {journal} {\bibinfo
  {journal} {Review of Scientific Instruments}\ }\textbf {\bibinfo {volume}
  {88}},\ \bibinfo {pages} {123507} (\bibinfo {year} {2017})}\BibitemShut
  {NoStop}%
\bibitem [{\citenamefont {Liu}\ \emph {et~al.}(2017)\citenamefont {Liu},
  \citenamefont {Hesse}, \citenamefont {Guo}, \citenamefont {Daughton},
  \citenamefont {Li}, \citenamefont {Cassak},\ and\ \citenamefont
  {Shay}}]{Liu2017}%
  \BibitemOpen
  \bibfield  {author} {\bibinfo {author} {\bibfnamefont {Y.}~\bibnamefont
  {Liu}}, \bibinfo {author} {\bibfnamefont {M.}~\bibnamefont {Hesse}}, \bibinfo
  {author} {\bibfnamefont {F.}~\bibnamefont {Guo}}, \bibinfo {author}
  {\bibfnamefont {W.}~\bibnamefont {Daughton}}, \bibinfo {author}
  {\bibfnamefont {H.}~\bibnamefont {Li}}, \bibinfo {author} {\bibfnamefont
  {P.~A.}\ \bibnamefont {Cassak}},\ and\ \bibinfo {author} {\bibfnamefont
  {M.~A.}\ \bibnamefont {Shay}},\ }\bibfield  {title} {\bibinfo {title} {Why
  does steady-state magnetic reconnection have a maximum local rate of order
  0.1?},\ }\href {https://doi.org/10.1103/PhysRevLett.118.085101} {\bibfield
  {journal} {\bibinfo  {journal} {\prl}\ }\textbf {\bibinfo {volume} {118}},\
  \bibinfo {pages} {085101} (\bibinfo {year} {2017})}\BibitemShut {NoStop}%
\bibitem [{Note1()}]{Note1}%
  \BibitemOpen
  \bibinfo {note} {In this configuration, the stationarity is not fulfilled,
  and thus the relation $E=E_r$ cannot be used.}\BibitemShut {Stop}%
\bibitem [{\citenamefont {{Smets, R.}}(2020)}]{heckle}%
  \BibitemOpen
  \bibfield  {author} {\bibinfo {author} {\bibnamefont {{Smets, R.}}},\
  }\href@noop {} {\bibinfo {title} {the heckle code}},\ \bibinfo {howpublished}
  {\url{https://github.com/rochSmets/heckle}} (\bibinfo {year}
  {2020})\BibitemShut {NoStop}%
\bibitem [{\citenamefont {Birdsall}\ and\ \citenamefont
  {Langdon}(2018)}]{Birdsall2018}%
  \BibitemOpen
  \bibfield  {author} {\bibinfo {author} {\bibfnamefont {C.}~\bibnamefont
  {Birdsall}}\ and\ \bibinfo {author} {\bibfnamefont {A.}~\bibnamefont
  {Langdon}},\ }\href {https://doi.org/10.1201/9781315275048} {\emph {\bibinfo
  {title} {Plasma Physics via Computer Simulation}}}\ (\bibinfo  {publisher}
  {{CRC} Press},\ \bibinfo {year} {2018})\BibitemShut {NoStop}%
\bibitem [{\citenamefont {Yao}\ \emph {et~al.}(2022)\citenamefont {Yao} \emph
  {et~al.}}]{Yao2022}%
  \BibitemOpen
  \bibfield  {author} {\bibinfo {author} {\bibfnamefont {W.}~\bibnamefont
  {Yao}} \emph {et~al.},\ }\bibfield  {title} {\bibinfo {title} {Detailed
  characterization of a laboratory magnetized supercritical collisionless shock
  and of the associated proton energization},\ }\href
  {https://doi.org/10.1063/5.0055071} {\bibfield  {journal} {\bibinfo
  {journal} {Matter and Radiation at Extremes}\ }\textbf {\bibinfo {volume}
  {7}},\ \bibinfo {pages} {014402} (\bibinfo {year} {2022})}\BibitemShut
  {NoStop}%
\end{thebibliography}%

\end{document}